\documentstyle[12pt,psfig,amssymb]{article}

\def\mathrm#1{{\rm #1}}
\def\mathbf#1{{\bf #1}}
\def\emph#1{{\em #1\/}}
\newcommand{\bx}{\mathbf{x}}
\newcommand{\bk}{\mathbf{k}}
\newcommand{\bP}{\mathbf{P}}
\newcommand{\bbk}{b_\mathbf{k}^{\vphantom{*}}}
\newcommand{\bbkc}{b_\mathbf{k}^*}
\newcommand{\e}{\mathrm{e}}
\newcommand{\Imag}{\mathop{\mathrm{Im}}\nolimits}

\begin{document}

\title{Use of singular classical solutions for calculation of
    multiparticle cross sections in field theory.}
\author{Bezrukov F.\footnote{\texttt{fedor@ms2.inr.ac.ru}}}
\date{}
\maketitle

\begin{abstract}
  A method of reducing the problem of the calculation of tree
  multiparticle cross sections in $\varphi^4$ theory to the solution
  of a singular classical Euclidean boundary value problem is
  introduced.  The solutions are obtained numerically
%?
% in spherical modes representation
  in terms of the decomposition in spherical harmonics, and the
  corresponding estimates of the tree cross sections at arbitrary
  energies are found.  Numerical analysis agrees with analytical
  results obtained earlier in the limiting cases of large and small
  energies.
\end{abstract}

\section{Introduction}
\label{sec:1}

The main theoretical tool in quantum field theory is presently the
perturbation theory.  Perturbative calculations provided the majority
of the experimentally checked results.  Therefore, limits of
applicability of perturbative calculations are of considerable
interest.  On the one hand, there exist processes related to complex
vacuum structure in gauge theories and nontrivial classical solutions
to field equations which cannot be described by perturbation theory.
On the other hand, even in the topologically trivial sector at
relatively low energies, processes which are described poorly by
perturbation theory also exist.

Perturbative calculations provide reliable results only in weakly
coupled models where the expansion parameter---dimensionless coupling
constant---is much smaller than unity.  But even in such theories
situations are possible in which other competing small (or large)
parameters exist.  A typical example is a process with large number
$n$ of particles in the final state ($n$ being of the order of the
inverse coupling constant $\lambda^{-1}$).

In conventional perturbation theory even above the topologically
trivial vacuum, the naive estimate of the amplitude gives the
factorial dependence $n!$ on the multiplicity of the final state. This
enhancement can in principle overcome the suppression due to the
powers of the coupling constant.  At the tree level, it is possible to
find an exact expression for the amplitude of creation of $n$ real
particles by one virtual particle in the theory with lagrangian
density
\begin{equation}
  \label{Action}
  {\cal L} = \frac{1}{2}(\partial_\mu\varphi)^2
  -\frac{1}{2}\varphi^2-\frac{\lambda}{4}\varphi^4
\end{equation}
(the mass is set equal to one) in special kinematics, namely, when all
particles have zero spatial momenta \cite{Voloshin:1991mz},
\begin{equation}
  \label{Atresh}
  A_{1\to n}^\mathrm{tree} =
  n!\left(\frac{\lambda}{8}\right)^\frac{n-1}{2}
  \;.
\end{equation}
This result points towards complete breakdown of the usual
perturbative calculations at
$n\gtrsim\lambda^{-1}$
%$n\stackrel{>}{\sim}\lambda^{-1}$
because it contradicts
unitarity of the theory.

Thus, some non-perturbative method is required for the calculation of
these cross sections.  The limit we are interested in is
\begin{equation}
  \label{lim}
  \lambda\to0 \;,\quad
  \lambda n=\mathrm{fixed} \;,\quad
  \varepsilon = \mathrm{fixed}
  \;,
\end{equation}
where $\varepsilon=(E-n)/n$ is the average kinetic energy of the
outgoing particles in the centre of mass frame.  Existing perturbative
calculations \cite{Libanov:1994ug,Libanov:1995gh} strongly suggest
that in this limit the total cross section has the exponential form,
\begin{equation}
  \label{sigmadef}
  \sigma_{1\to n} \sim
  \exp\left(\frac{1}{\lambda}F(\lambda n,\varepsilon)\right)
  \;.
\end{equation}
This form implies the semiclassical calculability of the cross
sections.  A method to obtain the exponent $F(\lambda n,\varepsilon)$
in all loops was formulated in ref.\ \cite{Son:1996wz}, which required
solution of a certain classical boundary problem in complex time.  At
small $\lambda n$ one needs only to solve purely Euclidean equations
with special boundary conditions.  In conventional perturbative
approach this limit corresponds to the contribution of tree graphs,
that gives the following dependence on $\lambda$:
\begin{equation}
  \label{Fdef}
  F_\mathrm{tree}(\lambda n,\varepsilon) =
  \lambda n\ln\left(\frac{\lambda n}{16}\right)-\lambda n
  +\lambda nf(\varepsilon)
  \;.
\end{equation}
Let us note that in the domain of its applicability, i.~e., at
$\lambda n\ll 1$, this dependence means the exponential suppression of
the cross section, at least if $f(\varepsilon)$ does not become
infinite.  But as $\lambda n$ increases, the function
$F_\mathrm{tree}(\lambda n,\varepsilon)$ becomes positive and this
suppression disappears.  Thus, in the latter case one has to take into
account loop corrections to $F(\lambda n,\varepsilon)$ which are of the
order $(\lambda n)^2$ and higher \cite{Libanov:1995gh}.

Making use of the technique developed in refs.\ 
\cite{Son:1996wz,Rubakov:1995hq} it is possible, at least in
principle, to find the only unknown function $f(\varepsilon)$ in
(\ref{Fdef}).

Even in the simplest case of small $\lambda n$ the calculation of the
exponent $F_\mathrm{tree}$ at all energies $\varepsilon$ is rather
complicated (the method of the calculation is described in
section~\ref{sec:2}).  The corresponding classical solution has
singularities on a three-dimensional surface in four-dimensional
Euclidean space (in the case of four dimensional theory).  This
surface depends on $\varepsilon$ and is determined in the course of
the calculation.  The Rayleigh--Ritz variational procedure enables one
to obtain the lower bound on $F_\mathrm{tree}$.  Computational
procedure and its numerical realization are described in
sections~\ref{sec:3} and~\ref{sec:4}.  The purpose of this paper is to
explore the possibilities for the actual calculation of the tree
exponent (\ref{Fdef}).  In section~\ref{sec:5} our results are
compared to known analytical results in the limiting cases.

\section{Singular solutions and tree cross sections.  General
  formalism}
\label{sec:2}

Let us describe the technique to obtain the exponent for tree cross
sections in the limit (\ref{lim})
\cite{Son:1996wz,Rubakov:1995hq,Libanov:1997obzor}.  We consider the
process of the decay of one virtual particle of energy $E$ and
momentum $\bP=0$ into $n$ real particles in the model with the
lagrangian density (\ref{Action}).  Let us write the matrix element
$\langle\beta|S\varphi|0\rangle$ in the coherent state representation
\cite{Berezin:2ndquant-eng,FaddeevSlavnov:FuncInt-eng,Tinyakov:1993dr},
in $(d+1)$--dimensional Minkowskian space-time ($\langle\beta|$ is a
coherent state)
\begin{equation}
  \label{cohint}
  \langle\beta|S\varphi|0\rangle = 
  \lim_{
    T_i\to-\infty \atop T_f\to+\infty
    }
  \int D\varphi D\varphi_i D\varphi_f\,\varphi(E,\bP)
  \e^{B_i(\varphi_i)+B_f(\beta^*,\varphi_f)+i\int\!{\cal L}\,d^{d+1}x}
  \;,
\end{equation}
where
\begin{displaymath}
  \varphi(E,\bP)=\int\!dtd^d\!x\varphi(t,{\bx}){\e}^{-iEt+i\bP\bx}
  \;,\quad
  \varphi_\bk(t)=\int\frac{d^d\!x}{(2\pi)^{d/2}}\varphi(t,\bx)
  \e^{-i\mathbf{kx}}
  \;,
\end{displaymath}
\begin{displaymath}
  \varphi_i(\bk)=\varphi_\bk(T_i)
  \;,\quad
  \varphi_f(\bk)=\varphi_\bk(T_f)
  \;,
\end{displaymath}
and the boundary terms are
\begin{eqnarray*}
  B_i(\varphi_i) &=&
  -\frac{1}{2}\int\!d^d\!k\,\omega_\bk\varphi_i(\bk)\varphi_i(-\bk)
  \;, \\
  B_f(\beta^*,\varphi_f) &=&
  -\frac{1}{2}\int\!d^d\!k\,\omega_\bk\varphi_f(\bk)\varphi_f(-\bk) \\
  && -\frac{1}{2}\int\!d^d\!k\,\beta_\bk^*\beta_{-\bk}^*\e^{2i\omega_\bk T_f}
  +\int\!d^d\!k\sqrt{2\omega_\bk}\e^{i\omega_\bk T_f}\beta_\bk^*\varphi_f(-\bk)
  \;,
\end{eqnarray*}
where $\omega_\bk=\sqrt{1+\bk^2}$.  At the tree level the integral
(\ref{cohint}) is determined by the value of integrand
taken at the saddle point.  The extremum conditions for the exponent
are the classical field equation
\begin{equation}
  \label{cleqns}
  \partial_\mu^2\varphi+\varphi+\lambda\varphi^3 = 0
\end{equation}
and the following boundary conditions:
\begin{equation}
  \label{4*}
  \varphi_\bk(t) \stackrel{t\to-\infty}{\longrightarrow}
  a_\bk\e^{i\omega_\bk t}
  \;,\quad
  \varphi_\bk(t) \stackrel{t\to+\infty}{\longrightarrow}
  \frac{\beta_\bk^*}{\sqrt{2\omega_\bk}}\e^{i\omega_\bk t}
  +c_\bk\e^{-i\omega_\bk t}
  \;,
\end{equation}
where $a_\bk$ and $c_\bk$ are arbitrary.  At $t\to-\infty$ the
solution $\varphi_c(\beta^*,t,\bx)$ has only positive frequency part.
Energy conservation then implies that at $t\to+\infty$ the
solution should not contain negative frequency parts either, i.~e.,
$c_\bk=0$, and the exponent in eq.\ (\ref{cohint}) is zero.  Thus, the
matrix element in the coherent state representation has the form of the
Fourier component of the saddle point solution,
\begin{equation}
  \label{ae}
  A_E(\beta^*) \equiv
  \langle\beta|S\varphi|0\rangle_\mathrm{tree} =
  \int\!dtd^d\!x\,\varphi_c(\beta,t,\bx)\e^{-iEt+i\bP\bx}
  \;.
\end{equation}
As follows from the coherent state formalism, the tree amplitude of
the process $1\to n$ can be obtained from the matrix element
(\ref{ae}) in the following way,
\begin{equation}
  \label{ampl}
  A_{1\to n}(\bk_1,\ldots,\bk_n) =
  \left.
    \frac{\partial^nA_E(\beta^*)}{
      \partial\beta^*_{\bk_1}\ldots\partial\beta^*_{\bk_n}}
  \right|_{\beta^*=0}
  \;.
\end{equation}

To find the $n$-particle cross section, let us introduce the generating 
function
\begin{equation}
  \label{crossgenf}
  \Sigma(\xi,E) = \frac{1}{Z}\int D\beta D\beta^*
  \exp\left\{-\int\!d^d\!k\,\beta^{\vphantom{*}}_\bk\beta_\bk^*\right\}
  A_E(\sqrt{\xi}\beta^*)\bar{A}_E(\sqrt{\xi}\beta)
  \;,
\end{equation}
where $Z$ is the normalization factor.  The total cross section is
then given by the following formula,
\begin{displaymath}
  \sigma^\mathrm{tree}_{1\to n}(E,n) =
  \left.
    \frac{1}{n!}\frac{\partial^n}{\partial\xi^n}\Sigma(\xi,E)
  \right|_{\xi=0}
  \;.
\end{displaymath}
One can check this relation by differentiating the right hand side of
eq.\ (\ref{crossgenf}) and using eq.\ (\ref{ampl})
\cite{Rubakov:1995hq,Rubakov:1992fb}.

Making use of the Cauchy formula one can rewrite the expression for
$\sigma^\mathrm{tree}_{1\to n}$ in the following form
\begin{equation}
  \label{5}
  \sigma_{1\to n}^\mathrm{tree} =
  \frac{1}{Z}\oint\frac{d\xi}{\xi^{n+1}}\int D\beta D\beta^*
  \exp\left\{-\frac{1}{\xi}\int d^dk\beta^*\beta\right\}
  A_E(\beta^*)\bar{A}_E(\beta)
  \;.
\end{equation}
This integral can be calculated again in the saddle point
approximation, after taking care of zero modes corresponding to time
translations and possible exponentially large factors in
$A_E(\beta^*)$.  To get rid of both of them, let us introduce the
following variables,
\begin{displaymath}
  \beta^*_\bk=b^*_\bk\e^{i\omega_\bk t_0-i\bk\bx_0}\;.
\end{displaymath}
In terms of these variables we have
\begin{displaymath}
  \begin{array}{rcl}
    \varphi_c(\beta^*,t,\bx) &=& \varphi(b^*,t+t_0,\bx+\bx_0) \;,\\
    A_E(\beta^*) &=& A_E(b^*)\e^{iEt_0-i\bP\bx_0} \;.
  \end{array}
\end{displaymath}
Here $t_0$, $\bx_0$ are collective coordinates and $b_\bk$ are new
integration variables, obeying some constraint that fixes
translational invariance.  The form of this constraint will be
determined later.  In terms of new
variables, eq.\ (\ref{5}) becomes
\begin{eqnarray}
  \nonumber
  \lefteqn{\sigma_{1\to n}^\mathrm{tree} =
    \frac{1}{Z}\oint\frac{d\xi}{\xi}\int Db Db^*
    dx_0 dx_0' J A_E(b^*)\bar{A}_E(b)\times} \\
  \label{5*}
  &&
  \times\exp\begin{array}[t]{@{}r@{}l}
    \left[ \vphantom{\frac{1}{\xi}\int} \right. &
    iE(t_0-t_0')-i\bP(\bx_0-\bx_0') \\
    & \left.
      -\frac{1}{\xi}\int\!d^d\!k\,\bbkc\bbk\e^{i\omega_\bk(t_0-t_0')
        -i\bk(\bx_0-\bx_0')}
      - n\ln\xi
    \right]
    \;,
  \end{array}
\end{eqnarray}
where $J$ contains $\delta$-function of the constraint on $b_\bk$ and
the corresponding Faddeev--Popov; the later that does not make
exponential contribution and will not be considered in what follows.
The integration over $(x_0+x_0')$ gives the volume factor canceling
out with $Z$.  If there are no more exponentially large factors in
$A_E$, we can use the saddle point approximation (the saddle point of
the variable $(\bx_0-\bx_0')$ is equal to zero because we work in the
centre of mass frame $\bP=0$; we will not write this variable later
on):
\begin{equation}
  \label{crosssec}
  \sigma^\mathrm{tree}(E,n) \propto \e^{W_\mathrm{tree}^\mathrm{extr}}
  \;,
\end{equation}
where $W_\mathrm{tree}^\mathrm{extr}$ is the extremum value of the
functional
\begin{equation}
  \label{W}
  W_\mathrm{tree}(T,\theta,\bbk,\bbkc)
  =ET-n\theta-\e^{-\theta}\int\!d^d\!k\,\bbkc\bbk\e^{\omega_\bk T}
\end{equation}
over $T=i(t_0-t_0')$, $\theta=\ln\xi$, $\bbk$ and $\bbkc$.

Let us now determine the constraint on $\bbk$.
It should break the translational invariance.  We have already
mentioned that we need the condition on $\bbk$ to get rid of the
exponential factors in $A_E(\bbk)$.  Let us continue analytically the
solution of eq.~(\ref{cleqns}) to the Euclidean time.  Then the
boundary condition (\ref{4*}) and the absence of negative frequency
parts, $c_\bk=0$, implies that the solution should decay at $\Imag
t=\tau\to+\infty$, where $\tau$ is the Euclidean time.  We are not
interested in instanton effects, i.~e., we do not consider classical
solutions regular in Euclidean space (in $\lambda\varphi^4$ theory
there are no such solutions if $\lambda>0$).  Then, $\varphi_c$ should
be singular somewhere in Euclidean space-time.  Generally, $\varphi_c$
is singular on some surface $\tau=\tau_s(\bx)$, where $\tau_s(\bx)<0$
for solutions smooth on the real time axis.  The behaviour of the
integral (\ref{ae}) is determined by singularities of the function
$\varphi_c$, i.~e., it is proportional to $\exp(E\tau_m+i\bP\bx_m)$,
where $\tau_m$ and $\bx_m$ are coordinates of the singularity closest
to the real axis ($\tau_m<0$).  Thus, to get rid of exponential
factors in $A_E(\beta^*)$ we need $\tau_m\to0$, $\bx_m=0$.  This
means, in other words, that we require the singularity surface in
Euclidean space-time to touch\footnote{Let us emphasize that we
  require only that $\tau_s$ \emph{approaches} zero ($\tau_s\to0$ at
  the point $\bx=0$).} the plane $\tau=0$ at the point $\bx=0$, i.~e.,
$\tau_s(\bx=0)=0$; $\tau_s(\bx)<0$ at $\bx\neq0$.  This condition
simultaneously fixes the translational invariance (in complex time),
so it is indeed a constraint fixing zero modes.

%It is useful to continue
%analytically the solution of eq.\ (\ref{cleqns}) to the Euclidean
%time, $t\to i\tau$.  In general, the solution decays as $\tau\to\infty$
%in view of eq.\ (\ref{4*}) and $c_\bk=0$; and it is singular on some surface
%$\tau=\tau_s(\bx)$.  We require that the solution exists at all
%values of real (Minkowskian) time, so we require $\tau_s(\bx)=-\Imag
%t<0$.  The leading exponential behaviour of the integral (\ref{ae}) is
%determined by the maximum value of $\tau_s(\bx)$.  If it is equal to
%$\tau^*$ then
%\begin{displaymath}
%  A_E \propto \e^{-E\tau^*}
%  \;.
%\end{displaymath}
%This implies that $A_E$ may indeed be regarded as a pre-exponential
%factor in eq.\ (\ref{5*}) provided that is $\tau^*\to0$.  This
%condition simultaneously fixes the translational invariance (in
%complex time).  Thus the constraint on $\bbk$ is the requirement that
%$\varphi_c(\tau,x)$ is singular at the point $\tau=0$, $\bx=0$.
%Furthermore, as we work in the centre-of-mass frame and there is no
%preferred spatial direction inherent in the problem ($\bP=0$), the
%singularity surface (and the solution itself) is symmetric under
%spatial rotations ($O(d)$-symmetry).  Different sets of $\bbk$,
%$\bbkc$ correspond to different singularity surfaces $\tau_s(\bx)$, so
%extremization over Fourier-components of the field and over the
%singularity surfaces are equivalent.

So, the problem of finding the tree cross sections at any $E$ and $n$
can be formulated in Euclidean space-time and consists of the
following steps:
\begin{itemize}
\item Find $O(d)$-symmetric solutions $\varphi(\tau,\bx)$ of the
  Euclidean field equations
  \begin{equation}
    \label{eeqns}
    \partial^2\varphi-\varphi-\lambda\varphi^3 = 0
    \;,
  \end{equation}
  which is singular on the surface $\tau_s(\bx)\le0$, $\tau_s(0)=0$
  and has the following asymptotics at $\tau\to\infty$:
  \begin{equation}
    \label{bk}
    \int\frac{d^d\!x}{(2\pi)^{d/2}}\varphi(\tau,\bx)\e^{-i\bk\bx} =
    \frac{\bbkc}{\sqrt{2\omega_\bk}}\e^{-\omega_\bk\tau}
    \;.
  \end{equation}
\item Calculate its frequency components $\bbk$ and determine $W$
  according to eq.\ (\ref{W}).
\item The functional $W$ should then be extremized over variables
  $\bbk$, $\bbkc$ (or, what is the same, over all singularity surfaces
  of the described type), $T$ and $\theta$.  The tree cross section of
  the process $1\to n$ is then given by the formula (\ref{crosssec}).
\end{itemize}

Analytical solutions of this boundary value problem can be found only
in special cases (they will be briefly described in
section~\ref{sec:5}).  Furthermore, in numerical computation it is
impossible to extremize the functional (\ref{W}) over infinite
dimensional space of singularity surfaces. One can only make use of
Rayleigh--Ritz procedure, i.~e., choose some finite dimensional
subclass of these surfaces and extremize the functional within this
subclass.  Let us consider this process more closely.  Let the
functional
\begin{math}
  \int\!d^d\!k\,\bbk\bbkc\e^{\omega_\bk T}
\end{math}
reach its minimal value for some $\bbk$:
\begin{math}
  \int\!d^d\!k\,\left.\bbk\bbkc\e^{\omega_\bk T}\right|_\mathrm{min}
  = C(T) > 0
\end{math}.
Let us fix some family of singularity surfaces
$\Sigma(T)$.  For these surfaces we have
\begin{math}
  \int\!d^d\!k\,\left.
    \bbk\bbkc\e^{\omega_\bk T}
  \right|_\mathrm{\Sigma(T)}
  = C_\Sigma(T) \ge C(T)
\end{math}
for all $T$.  After inserting the saddle point value for $\theta$,
equal to $\theta=-\ln n+\ln C(T)$, we get
\begin{eqnarray*}
  W(T)=n\ln n-n+ET-n\ln C(T) && \mbox{has an extremum at $T_1$,} \\
  W_\Sigma(T)=n\ln n-n+ET-n\ln C_\Sigma(T) &&
  \mbox{has an extremum at $T_2$.}
\end{eqnarray*}
Comparing $W(T_1)$ and $W_\Sigma(T_2)$ one can obtain the following
inequalities:
\begin{eqnarray*}
  W(T_1) \ge W(T_2) \ge W_\Sigma(T_2) &&
  \mbox{if $W$ has a maximum at $T_1$;} \\
  W_\Sigma(T_2) \le W_\Sigma(T_1) \le W(T_1) &&
  \mbox{if $W_\Sigma$ has a minimum at $T_2$}
\end{eqnarray*}
(in the real computation the second case is realized).  Thus if we
limit ourselves to some subclass of singularity surfaces, we get a
lower bound on the exact value of $W_\mathrm{tree}(E,n)$.

\section{Expansion in spherical modes}
\label{sec:3}

The following calculations will be done in $(3+1)$-dimensional
space-time.  We will consider only compact singularity surfaces.

The only requirement imposed on the singularity surface is that it
touches the plane $\tau=0$ at the point $\bx=0$ and $\tau_s(\bx)<0$ at
all other spatial points.  So, we can describe the singularity surface
using the following method.  Let us choose a sphere of the radius
$R_s$ with the centre at the origin of the coordinate system.  Let the
field configuration $\varphi$ be infinite at the point $\tau=R_s$,
$\bx=0$ and finite at all points $\sqrt{\bx^2+\tau^2}>R_s$.  Then the
singularity surface for this field touches the plane $\tau=R_s$ at
$\bx=0$ and is contained inside the chosen sphere, i.~e.\ at
$\tau_s(\bx)\le R_s$.  This description is suitable for singularity
surfaces of the form of a sphere that is slightly squeezed along the
horizontal direction.  These configurations are ones of primary
interest, as we will see from the results of the calculations.  We
should only make the substitution
\begin{math}
  \tau \to \tau+R_s
  \;,
\end{math}
to move the singularity to the origin.  This is equivalent to the
following change in frequency components of the field,
\begin{equation}
  \label{bkshift}
  \bbk = \tilde\bbk\e^{-\omega_\bk R_s}
  \;,
\end{equation}
where $\tilde\bbk$ are Fourier components of the field singular at the
point $(R_s,0)$.

As far as the field configuration is $O(3)$ symmetric, the field is a
function of two variables, $\varphi(\rho,\theta)$, where $\theta$ is
the angle between the radius-vector and $\tau$ axis, and $\rho$ is the
length of the radius-vector (in 4-dimensional Euclidean space).  The
Euclidean field equations can be obtained by varying the following
action:
\begin{equation}
  \label{Spolar}
  S = 4\pi\int\limits_0^\pi\! d\theta\!
  \int\limits_{\lefteqn{\scriptstyle\rho_\mathrm{min}(\theta)}}^\infty
  \!d\rho\,
  \rho^3\sin^2\theta\left\{
    \frac{1}{2}\left(\frac{\partial\varphi}{\partial\rho}\right)^2
    +\frac{1}{2\rho^2}\left(\frac{\partial\varphi}{\partial\theta}\right)^2
    +\frac{1}{2}\varphi^2+\frac{\lambda}{4}\varphi^4
  \right\}
  \;.
\end{equation}
Let us make use of the expansion in spherical modes,
\begin{equation}
  \label{GegModes}
  \varphi(\rho,\theta) =
  \sum\limits_{n=0}^\infty
  \varphi_n(\rho)C_n^{(1)}(\cos\theta)
  \;,
\end{equation}
where $C_n^{(1)}(\cos\theta)=\frac{\sin(n+1)\theta}{\sin\theta}$ are
Gegenbauer polynomials.
Asymptotically, as $\rho\to\infty$, the functions
$\varphi_n(\rho)$ are of the form
\begin{equation}
  \label{assimpt}
  \varphi_n(\rho) = a_n \frac{K_{n+1}(\rho)}{\rho}
  \;,
\end{equation}
where $K_n$ are
modified Bessel
%McDonald
functions.  The coefficients $\tilde\bbk$ of the
expansion of this field configuration in plain waves are
\begin{displaymath}
  \tilde\bbk = \sum_{n=0}^\infty a_n
  \sqrt{2\pi}\frac{C_n^{(1)}(\omega_\bk)}{\sqrt{2\omega_\bk}}
  \;,
\end{displaymath}
so the integral in eq.\ (\ref{W}) can be expressed in terms of the
coefficients $a_n$ as follows,
\begin{equation}
  \label{Ibkbk}
  I(z) \equiv \int\tilde\bbkc\tilde\bbk\e^{-\omega_\bk z} d^3\bk
  = 2\pi^2\sum\limits_{n,m=0}^\infty a_na_m
  \Big[K_{n+m+2}(z)-K_{n-m}(z)\Big]
\end{equation}
Upon substituting the mode expansion (\ref{GegModes}) into eq.\ 
(\ref{Spolar}), we get the expression for the action in terms of the
spherical modes.  Its extremization yields the equation (\ref{eeqns})
in terms of radial functions $\varphi_n(\rho)$.

The radial functions should have the form (\ref{assimpt}) at
$\rho\to\infty$, i.~e., they should not have growing components.

One also has to impose the second boundary condition which will ensure
that the field becomes infinite on some singularity surface, subject
to all requirements mentioned in the beginning of this section.  To
formulate it precisely, one has to move a bit away from the
singularity, i.~e., the condition $\varphi(R_s,0)=\infty$ should be
substituted by the condition $\varphi(R,0)=A$, where
$A\gg1/\sqrt{\lambda}$.  In this case one can neglect the mass term in
the field equation near the point $(R_s,0)$ and approximate the
singularity surface by a plane.  Then $\varphi$ in this region is
\begin{displaymath}
  \varphi = \sqrt{\frac{2}{\lambda}}\frac{1}{l(x)}
  \;,
\end{displaymath}
where $l(x)$ is the distance from the point $x$ to the singularity
surface.  It is straightforward to see that the singularity is
placed at the following distance from the origin,
\begin{equation}
  \label{nsing}
  R_s = R-\sqrt{\frac{2}{\lambda}}\frac{1}{A} \;.
\end{equation}

Thus, the singularity surface satisfying the required constraints (its
form will be described in more detail later) is determined by a set of
spherical components $c_n = \varphi_n(R)$, which should satisfy the
condition
\begin{equation}
  \label{constr1}
  \sum\limits_{n=0}^\infty c_n(n+1) = A
  \;,
\end{equation}
i.~e., $\varphi(R,\theta=0)=A$; and also the condition
\begin{equation}
  \label{constr2}
  \varphi(R,\theta\neq0)\le A
  \;,
\end{equation}
which in the simplest case of two non-zero components $c_n$ is
reduced to the requirement that both are non-negative.

The simplest configurations are $O(4)$ symmetric.  They are defined in
such a way that $c_0=A$ and $c_n=0$ for all other $n$, and are
characterized by only one parameter, the radius of the
singularity surface $R$.

Let us extremize over the parameters $T$, $\theta$ and singularity
surfaces.  Making use of eqs.\ (\ref{bkshift}) and (\ref{Ibkbk}),
one writes the expression (\ref{W}) in the following form,
\begin{displaymath}
  W_\mathrm{tree}(T,\theta,\bbk,\bbkc)
  =ET-n\theta-I(z)\e^{-\theta}
  \;,
\end{displaymath}
where $z=2R_s-T$.  The stationarity conditions for this expression over
$T$ and $\theta$ can be easily obtained,
\begin{equation}
  \label{Eextr}
  n = I(z)\e^{-\theta} \;,\qquad
  E = -I'(z)\e^{-\theta}
  \;,
\end{equation}
where $I'(z)$ is the derivative of the expression (\ref{Ibkbk}).
Finally,
\begin{eqnarray}
  \label{Wf}
  W_\mathrm{tree}(\bbk,\bbkc) &=&
  n\ln\frac{\lambda n}{16}-n+nf(\varepsilon)
  \;, \\
  \label{fe}
  f(\varepsilon) &=& (\varepsilon+1)T+\ln16-\ln\lambda I(z)
  \;,
\end{eqnarray}
where $T$ should be expressed through $\varepsilon$ by solving the
following equation,
\begin{equation}
  \label{et}
  \varepsilon+1 = -\frac{I'(z)}{I(z)}
  \;,
\end{equation}
which is a consequence of eq.\ (\ref{Eextr}).

For calculational reasons it is more convenient to perform the
extremization in a slightly different order: first fix some value of
$T$, then find the minimum of $I(z)$ over all singularity surfaces
($\bbk$ and $\bbkc$) and, finally, obtain the corresponding value of
$\varepsilon$ from eq.~(\ref{et}).

Let us note that the function $f(\varepsilon)$ is independent of
$\lambda$.  Indeed, let us make the substitution
\begin{displaymath}
  \varphi \to \frac{\tilde\varphi}{\sqrt{\lambda}}
  \;.
\end{displaymath}
Then eq.\ (\ref{eeqns}) becomes the equation for $\tilde\varphi$
with $\lambda=1$, and the integral (\ref{Ibkbk}) transforms into
\begin{displaymath}
  I(z) \to \frac{\tilde{I}(z)}{\lambda}
  \;.
\end{displaymath}
The dependence on $\lambda$ in eq.\ (\ref{fe}) disappears.  Thus one
can set $\lambda$ equal to one for the calculation of
$f(\varepsilon)$.

\begin{figure}[t]
  \begin{center}
    \leavevmode
    \psfig{file=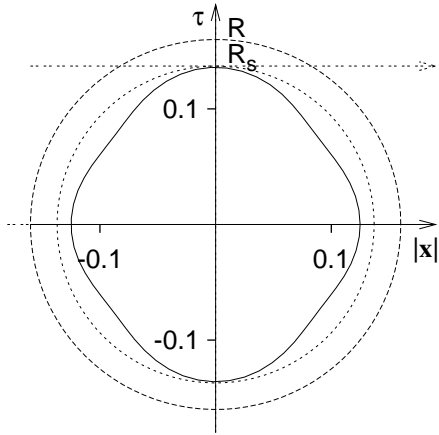}
    \caption{Singularity surface at energy $\varepsilon=10$
      (solid line).  The 4th harmonic is turned on at infinity.  A
      sphere of the radius $R$, on which one can determine $c_n$, is
      also shown.}
    \label{fig:singreal}
  \end{center}
\end{figure}

In the case when ${c_n\ll c_0}$ (or, equivalently, ${c_n\ll A}$) for
all ${n>0}$, it is possible to determine the deviation of the
singularity surface of the corresponding field configuration from
sphere.  In this case the field $\varphi$ is large at $\rho=R$ for all
$\theta$, so in this region we can use the approximation of massless
field.  We will also assume that the radius $R$ is large
enough to consider the singularity surface flat at all points.  Then,
by making use of eq.\ (\ref{nsing}), we immediately get
\begin{displaymath}
  \Delta R_s(\theta) = \sqrt{\frac{2}{\lambda}}\left(
    \frac{1}{\varphi(R,\theta)}-\frac{1}{\varphi(R,0)}
  \right)
  \;,
\end{displaymath}
where $\Delta R_s(\theta)=R_s(0)-R_s(\theta)$ characterizes the
deviation of the singularity surface from sphere.  The shape of a
typical singularity surface found from our calculations is shown in
fig.~\ref{fig:singreal}.

\section{Numerical calculation of tree cross sections}
\label{sec:4}

\begin{figure}[t]
  \begin{center}
    \leavevmode
    \psfig{file=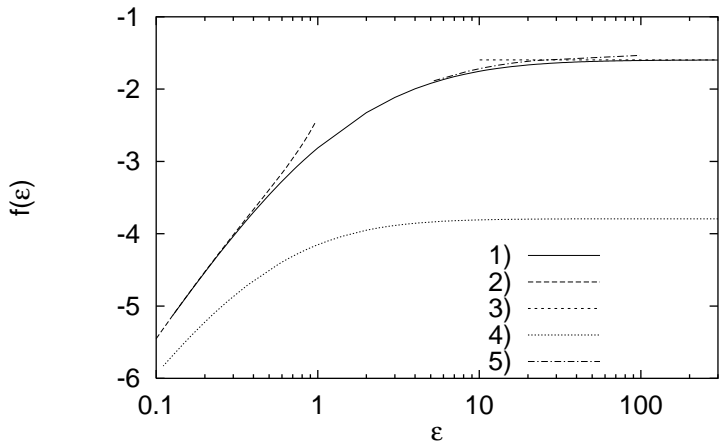}
    \caption{Bounds on $f(\varepsilon)$.  1) $O(4)$-symmetric bound.
      2) Analytical estimate for small energies $\varepsilon$.  3)
      Analytical lower bound for high energies.  4) Voloshin's lower
      bound.  5) The bound obtained with the 4th harmonic turned on at
      infinity.}
    \label{fig:2}
  \end{center}
\end{figure}

In the case of $O(4)$-symmetric solutions (one-parameter family of
singularity surfaces) the problem is simple: it is reduced to one
ordinary differential equation for $\varphi_0(\rho)$.  It is even not
necessary to solve the boundary value problem, one can merely fix
different values of $a_0$ (i.~e., initial conditions at infinite
$\rho$) and find the corresponding singularity radii $R_s$.  The lower
bound on $f(\varepsilon)$ obtained in this way is shown in
fig.~\ref{fig:2} by a solid line.

\begin{figure}[t]
  \begin{center}
    \leavevmode
    \psfig{file=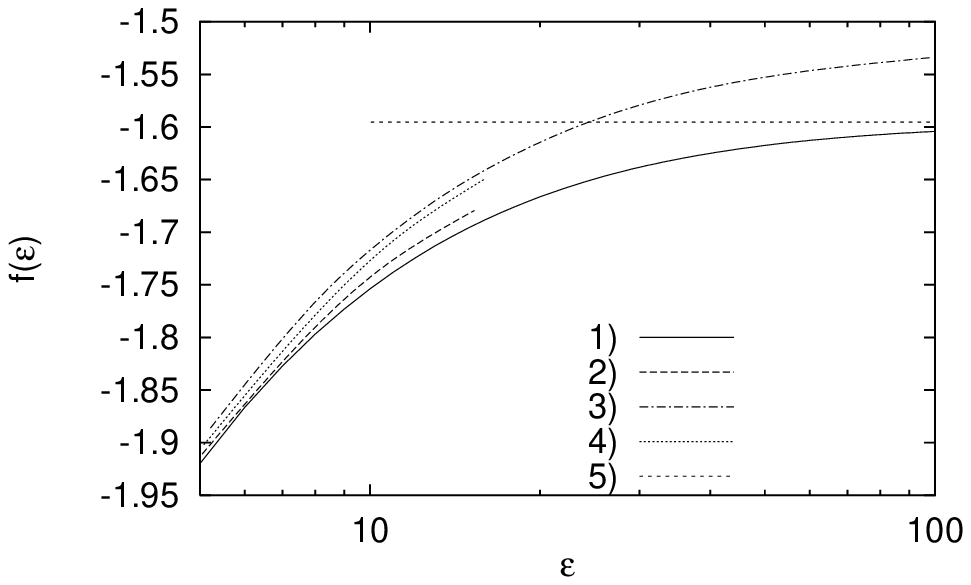}
    \caption{Bounds on $f(\varepsilon)$
      (scaled part of fig.~\protect\ref{fig:2}).  1) $O(4)$-symmetric
      bound.  2) Bound obtained with the 2nd harmonic turned on at
      infinity.  3) 4th harmonic. 4) 6th harmonic.  5) Analytical
      lower bound for high energies.}
    \label{fig:2scaled}
  \end{center}
\end{figure}

If we do not limit ourselves to spherically symmetric modes only, we
should, in principle, solve the boundary value problem (\ref{eeqns}).
It is difficult to solve it directly because one has to find
$\varphi_n$ that rapidly decrease with the harmonic number $n$ to have
the sum in eq.~(\ref{Ibkbk}) convergent (and this is required at large
$\rho$, where the field itself is small).  Furthermore, in direct
approach the configuration is determined by the shape of the
singularity surface, where the values of the field are large, so the
problem is extremely unstable.  For the same reason one is unable to
solve the boundary value problem formulated in terms of spherical
components by fixing $\varphi_n$ at the radius $R$ near the
singularity.

However, one can invent a method that is similar to one used for the
$O(4)$-symmetric problem.  Let us fix the values of $a_n$, i.~e.,
{}spherical modes at infinity (that is equivalent to fixing $\bbk$)
and solve the system of \emph{ordinary} differential equations with
\emph{initial} conditions specified in this way.  At some finite
radius $R$ the field becomes infinite (more precisely, becomes larger
than some prescribed large number $A$).  If the coefficients
$c_n=\varphi_n(R)$ at this $R$ obey the conditions (\ref{constr1}) and
(\ref{constr2}), then this configuration satisfies all requirements
mentioned above.  We found that this property is satisfied in a
sufficiently large region of parameters $a_n$, i.~e., radial functions
at infinity.  In particular, one can set all $a_n$ equal to zero for
all $n\neq0,k$ for any fixed $k$.

The major problem of this method (and presumably any method that makes
use of the expansion in frequency components) is that the number of
spherical modes used for computation is limited.  However, the errors
caused by this truncation are significant only at short distances from
the singularity (in the region of strong non-linearity) and do not
lead to a considerable error in the determination of $R_s$, at least
for not too high energies.

To search for the minimum of the $I(2R_s-T)$ as a function of $a_n$
(that effectively means the minimization over $\bbk$, $\bbkc$) we used
the multidimensional downhill simplex method \cite{num_recipes}, that
does not require any additional information about the function which
is extremized (i.~e.\ it does not need its derivatives).  To find the
minimum, about 50 calculations of $I(2R_s-T)$ had to be made for
different configurations.

The calculations were performed with turning on two non-zero spherical
modes at infinity, $a_0$ and $a_k$.  All values of $k$ smaller than
$8$ were explored.  The maximum deviation from the $O(4)$--symmetric
result was obtained at $k=4$.  Let us recall in this respect that we
obtain the lower bound on $f(\varepsilon)$ by the Rayleigh--Ritz
procedure, so we are interested in the maximum value of
$f(\varepsilon)$.  The results for $k=4$, $2$ and $6$ are shown in the
fig.~\ref{fig:2scaled}.  At smaller energies $\varepsilon$ the
difference from the spherical calculation is negligible while at
larger $\varepsilon$ the singularity radius $R_s$ becomes small and
relative error in its calculation grows.  The $O(4)$--symmetric
calculations were performed in a wider interval of energies, which is
shown in fig.~\ref{fig:2} together with some analytical results
(see section~\ref{sec:5}).

The shape of the singularity surface corresponding to the minimal
value of $I(2R_s-T)$ is also of some interest.  The singularity
surface for $\varepsilon=10$ and $k=4$ is shown in
fig.~\ref{fig:singreal}.

\section{Comparison of analytical and numerical results}
\label{sec:5}

In the limiting cases of small and large energies it is possible to
implement the procedure of the section~\ref{sec:2} analytically
\cite{Son:1996wz,Bezrukov:1995qh}.  Let us present the analytical
results here for comparison.

In case of small energy $\varepsilon$ one may start with the
solutions of the following form,
\begin{displaymath}
  \varphi(\tau,\bx) = \sqrt{\frac{2}{\lambda}}
  \frac{1}{\sinh(\tau-\tau_s(\bx))}
  \;,
\end{displaymath}
which satisfies eq.~(\ref{eeqns}) up to the terms of order
$O((\partial_\bx\tau_s)^2)$ (the case when $\tau_s(\bx)=0$ for all
$\bx$ corresponds to the creation of particles at the threshold).  One
can also find corrections of the orders $(\partial_\bx\tau_s)^2$ and
$(\partial_\bx\tau_s)^4$ to this expression.  As a result, one obtains
the following estimate \cite{Son:1996wz,Bezrukov:1995qh},
\begin{equation}
  f(\varepsilon) = \frac{3}{2}\left(\ln\frac{\varepsilon}{3\pi}+1\right)
  -\varepsilon\frac{17}{12}
  +\frac{\varepsilon^2}{432}(1327-96\pi^2)+O(\varepsilon^3)
  \;. \label{9*'}
\end{equation}
This analytical result is shown in fig.~\ref{fig:2} by a dashed line; it
coincides with our numerical result at $\varepsilon<0.5$.

In the case of high energies one may try to neglect the mass term in
the field equation and consider the massless $\varphi^4$ theory.  In
massless theory, an $O(4)$-symmetric solution---the Fubini--Lipatov
instanton---is known \cite{Fubini:1976jm,Lipatov:1977hj}.  This
solution may be used to construct the solution which is singular at
the point $\tau=\bx=0$ \cite{Khlebnikov:1992af}:
\begin{equation}
  \label{Khl}
  \varphi_0 =
  \sqrt{8\over\lambda}
  {R_s\over\bx^2+(\tau+R_s)^2-R_s^2}
  \;,
\end{equation}
where $R_s$ is the collective coordinate determining the size of the
singularity surface.  From this solution one obtains the following
bound \cite{Son:1996wz,Khlebnikov:1992af},
\begin{equation}
  \label{fultrarel}
  f(\varepsilon\to\infty) \ge \ln\frac{2}{\pi^2}
  \;.
\end{equation}
This bound is consistent with our numerical calculations at
$\varepsilon>50$.

It is of interest to compare our numerical results with other existing
estimates.  An alternative lower bound on $f(\varepsilon)$, following
from the direct analysis of Feynman diagrams, is given in
ref.~\cite{Voloshin:1992rr}.  As is clear from fig.~\ref{fig:2} our
bound is considerably more stringent.

\section{Conclusion}

In this paper we described a method for the calculation of
multiparticle cross sections in $\varphi^4$ theory at the tree level.
Our numerical results obtained in the $O(4)$--symmetric case coincide
with known limiting cases in the domains of validity of the latter.
More precise results obtained with larger family of singularity
surfaces indicate that even the simplest $O(4)$--symmetric
approximation gives nearly an exact answer.  The lower bounds obtained
in this paper are better than previous bounds.

This work is supported in part by Russian Foundation for Basic
Research, grant No.\ 96--02--17449a, and Soros Students Programme.
The author is grateful to V.~A.~Rubakov for numerous discussions at
all stages of the work and A.~N.~Kuznetsov, M.~V.~Libanov,
P.~G.~Tinyakov, and S.~V.~Troitsky for valuable remarks.

\nocite{Bezrukov:1995ta}
%\bibliographystyle{unsrt}
%\bibliography{index}

\end{document}